\documentclass[a4paper,10pt]{article}

\usepackage[T1]{fontenc}
\usepackage{lmodern}
\usepackage[utf8]{inputenc}
\usepackage[margin=1.0in]{geometry}
\usepackage[inline]{showlabels}
\usepackage{amsmath,amsfonts,amssymb,mathtools,mathrsfs,dsfont}
\usepackage{xcolor}
\usepackage{hyperref,url} 
\usepackage{graphicx}
\usepackage{subfig}
\usepackage{tikz}
\usepackage{comment}

\hypersetup{
  colorlinks,
  citecolor=violet,
  linkcolor=blue,
  urlcolor=blue}

\numberwithin{equation}{section}

\DeclareMathAlphabet\mathbfcal{OMS}{cmsy}{b}{n}

\newcommand{\beq}{\begin{equation}}
\newcommand{\eeq}{\end{equation}}

\newcommand{\Er}{{\mathbfcal{E}}}
\newcommand{\p}{\partial}

\begin{document}

\begin{titlepage}

\begin{flushright}
IFUM-1093-FT \\
LIFT-1-1.22
\end{flushright}

\vspace{2cm}

\begin{center}
\renewcommand{\thefootnote}{\fnsymbol{footnote}}
{\Huge \bf Black holes in a swirling universe}
\vskip 32mm
{\large {Marco Astorino$^{a,b}$\footnote{marco.astorino@gmail.com},
Riccardo Martelli$^{c}$\footnote{riccardo.martelli@studenti.unimi.it}
and Adriano Vigan\`o$^{b,c}$\footnote{adriano.vigano@unimi.it} }}\\

\renewcommand{\thefootnote}{\arabic{footnote}}
\setcounter{footnote}{0}
\vskip 10mm
\vspace{0.2 cm}
{\small \textit{$^{a}$Laboratorio Italiano di Fisica Teorica (LIFT),  \\
Via Archimede 20, I-20129 Milano, Italy}\\
} \vspace{0.2 cm}
{\small \textit{$^{b}$Istituto Nazionale di Fisica Nucleare (INFN), Sezione di Milano \\
Via Celoria 16, I-20133 Milano, Italy}\\
} \vspace{0.2 cm}
{\small \textit{$^{c}$Universit\`a degli Studi di Milano}} \\
{\small {\it Via Celoria 16, I-20133 Milano, Italy}\\
}
\end{center}
\vspace{5.4 cm}

\begin{center}
{\bf Abstract}
\end{center}
{

We present a new solution in Einstein's General Relativity representing a Schwarzschild black hole immersed in a rotating universe.
Such a solution is constructed analytically by means of the last unexplored Lie point symmetry of the Ernst equations for stationary and axisymmetric spacetimes.
This kind of Ehlers transformation is able to embed any given solution into a rotating background, which is not of Newman--Unti--Tamburino (NUT) type.
We analyse the physical properties, ergoregions and geodesics of the new metric, which is regular outside the event horizon and has a well defined thermodynamics.
We finally consider the Kerr generalisation.}

\end{titlepage}

\addtocounter{page}{1}

\section{Introduction}

The usefulness of solution generating techniques in General Relativity is renowned.
They allow one to build basically any solution of the vacuum theory without integrating the equations of motion, as conjectured by the Geroch~\cite{geroch1,geroch2}, and proved by Hauser and Ernst~\cite{hauser-ernst}.
Such techniques are fundamental to build rotating multi-black hole systems, or black rings and black Saturns in higher-dimensional gravity.
However, there are still unexplored aspects even in the oldest and basic generating methods.

In this article we want to deepen the study of the less known Lie point symmetries of the Ernst equations~\cite{ernst1,ernst2}, which are a smart incarnation of Einstein(--Maxwell) equations for stationary and axisymmetric spacetimes.
It is well known that the Ernst equations are invariant under the Ehlers and the Harrison transformations.
Such transformations are able to generate some of the most significant solutions of the theory.
For instance, starting from the Schwarzschild seed metric, these transformations lead to the Taub--NUT and the Reissner--Nordstr\"om black hole respectively, as shown in~\cite{reina-treves,ehlers-marcoa,charged-binary}.
It is also well known that Ernst himself was able to add an external electromagnetic field to black hole spacetimes trough a ``magnetic'' version of the Harrison transformation, see~\cite{ernst-magnetic,kerr-magnetic} and \cite{ernst-remove} for the static, the rotating and the accelerating cases.

For these reasons, it seems natural to ask oneself what is the effect of the ``magnetic'' version of the Ehlers transformation, which seems the only Lie point symmetry of the gravitational equations which remains to be analysed in details, up to the authors knowledge.
As mentioned above the other symmetries give rise to physical interesting spacetimes, therefore it is reasonable to presume that this last uncharted one might do so, perhaps it can generate even a metric before unknown in the literature.
It turns out that such a transformation embeds any given stationary and axisymmetric seed spacetime into a rotating background, which we will dub ``swirling universe'', for its peculiar characteristic. Indeed, the background can be interpreted as a gravitational whirlpool, and its frame dragging turns a static seed solution into a stationary metric.

In Sec.~\ref{sec:generation} we derive and present the transformation of the Ernst equation we will use.
In Sec.~\ref{sec:analysis} we test the consequences of the transformation on the spherical symmetric black hole, which will be our seed metric, as done with the symmetries previously analysed in the literature.
In this way we generate a novel and analytic exact solution of the Einstein equations, which generalises and deforms the Schwarzschild spacetime.
Of course, because of the well-known no-hair theorems for black hole in four-dimensional General Relativity the new solution could be a black hole only by renouncing to asymptotic flatness, in a similar way the black holes embedded in the external electromagnetic field of Melvin universe are \cite{melvin,ernst-magnetic}.
More precisely, we study the background spacetime in Sec.~\ref{sec:analysis-background}, while we analyse the full black hole spacetime in Sec.~\ref{sec:analysis-blackhole}, by studying the physical properties, the geodesics and the thermodynamics  of the novel solution.
We also briefly present the generalisation to the Kerr black hole in the rotating background in Sec.~\ref{sec:kerr}.
We conclude the paper by establishing a connection (via a double-Wick rotation) between our background metric and the Taub--NUT spacetime in Sec.~\ref{sec:double-wick}. 

\section{Generation of the solution}
\label{sec:generation}

It is well known~\cite{ernst1} that the vacuum Einstein equations $R_{\mu\nu}=0$ for stationary and axisymmetric spacetimes are equivalent to the Ernst equations
\begin{equation}
\label{ee-ernst}
(\text{Re}\ \Er) \, \nabla^2 \Er =
\vec{\nabla} \Er \cdot \vec{\nabla} \Er \,,
\end{equation}
where $\Vec{\nabla}$ and the various vectorial quantities are understood in Euclidean space with cylindrical coordinates $(\rho,\varphi,z)$.
The gravitational Ernst potential $\Er \coloneqq f + i h $ is a complex scalar function built from the characteristic functions of the general stationary and axisymmetric spacetime in pure General Relativity, represented by the Lewis--Weyl--Papapetrou (LWP) metric
\begin{equation} \label{LWP-e}
{ds}^2 = - f ( dt - \omega d\varphi )^2 + f^{-1} \bigl[ \rho^2 {d\varphi}^2 + e^{2\gamma}  \bigl( {d \rho}^2 + {d z}^2 \bigr) \bigr] \,.
\end{equation}  
These spacetimes, because of the requested symmetries, enjoy at least two commuting Killing vectors ($\partial_t$, $\partial_\varphi$) and three arbitrary functions which depend on the non-Killing coordinates $(\rho,z)$.
But, thanks to the integrability properties of the system, $\gamma(\rho,z)$ decouples from the other functions and is uniquely determined by $f(\rho,z)$ and $\omega(\rho,z)$, see~\cite{ehlers-marcoa} for a brief review of the Ernst formalism.
In order to take advantage of the integrability properties of the model, Ernst found out that it is convenient to define the imaginary part of the potential $\Er$ as follows
\begin{equation}
\label{h}
\vec{\nabla}   h \coloneqq
\frac{f(\rho,z)^2}{\rho} \vec{e}_\varphi \times \vec{\nabla} \omega \,.
\end{equation}
The remarkable feature of the Ernst technique consists in revealing non-trivial symmetries of the complex equations and the integrability of the system of coupled partial differential equations governing the model.
Note that the Ernst formulation of the gravitational Einstein equations is not just a redefinition of the function fields, but it is essential to generate physically non-trivial transformations, i.e.~transformations whose effect cannot be reabsorbed into the gauge freedom of the theory.
Therefore, they generate inequivalent metrics with respect to the starting seed.

We will mainly be interested in Lie point symmetries of the Ernst equation~\eqref{ee-ernst}, i.e.~symmetries associated to a continuous parameter (real or complex), which labels the symmetry transformation.
Obviously, Lie point symmetries do not deplete the whole set of symmetries of the model under consideration:
for instance, also discrete symmetries might be present.
In particular, in this article we will mainly deal with the Ehlers transformation, more details about the symmetry transformation of the Ernst equations can be found in~\cite{stephani,ehlers-marcoa,tesi-riccardo}.
In fact it is easy to verify that the following transformation 
\begin{equation}
\label{Ehlers-e}
E \, : \,  \Er \to \Er' = \frac{\Er}{1+i\jmath\Er}   \,, \end{equation}
with $\jmath \in {\mathbb R}$, leaves the Ernst equation~\eqref{ee-ernst} formally unchanged.
As stated in the introduction, when the above Ehlers transformation~\eqref{Ehlers-e} is applied to the Schwarzschild black hole, cast into the LWP form~\eqref{LWP-e}, one generates the Taub--NUT spacetime.
However, there is a subtlety here that we can take advantage of, in the very same way Ernst did, dealing with the Harrison transformation, for generating the Schwarzschild black hole embedded into the Melvin electromagnetic universe instead of the Reissner--Nordstr\"om metric.

There exist two non-equivalent forms of the LWP metric that can be used to construct the Ernst equations (\ref{ee-ernst}).
One form is the one given in~\eqref{LWP-e}, the other one is what we will dub the \emph{magnetic} LWP:
\begin{equation}
\label{LWP-m}
{d\bar{s}}^2 = \bar{f} ( d\phi - \bar{\omega} d\tau )^2 + \bar{f}^{-1} \bigl[ -\rho^2 d\tau^2 + e^{2\bar{\gamma}}  \bigl( {d \rho}^2 + {d z}^2 \bigr) \bigr] \,,
\end{equation} 
which is obtained from~\eqref{LWP-e} by a discrete transformation of the form
\begin{equation}
\label{W}
W \coloneqq
\biggl\{
f \to \frac{\rho^2}{\bar{f}} -\bar{f} \bar{\omega}^2 \,, \quad
\omega \to \frac{\bar{f^2} \bar{\omega}}{\bar{f}^2 \bar{\omega}^2-\rho^2} \,, \quad
e^{2\gamma} \to e^{2\bar{\gamma}}\left( \frac{\rho^2}{\bar{f}^2} -\bar{\omega}^2 \right) \biggr\} \,.
\end{equation}
Note that the above transformation, known as \emph{conjugation}~\cite{chandra-book}, acts as an involution operator, i.e.~$W \circ W = \mathds{1}$.
Just to give an example, we recall that equation~\eqref{W} maps the Schwarzschild black hole into the Witten bubble of nothing~\cite{horowitz}.
In fact the $W$ map can also be seen as an analytical continuation, or double-Wick rotation, of the non-Killing coordinates of the metric, as
\begin{equation}
\varphi \to  i \tau \,, \quad
t \to i \phi \,.
\end{equation}
The subtlety consists in casting our seed metric in the magnetic LWP form (\ref{LWP-m}) instead on the electric version (\ref{LWP-e}), otherwise it is well known we would just add the NUT parameter to the chosen seed \cite{reina-treves}, \cite{ehlers-marcoa}. At this point we have the full theoretical setting needed to proceed with the generation of a new solution.

First of all we have to choose the seed.
We start with the Schwarzschild black hole whose metric, in spherical coordinates, is
\begin{equation}
\label{schw}
{ds}^2 = - \biggl(1- \frac{2m}{r} \biggr) {d\tau}^2 + \frac{{dr}^2}{1- \frac{2m}{r}} + r^2 {d\theta}^2 + r^2 \sin^2 \theta {d \phi}^2 \,.
\end{equation}
The most convenient coordinates for the generating methods, in this case, are the spherical ones $(r,\theta)$, related to the Weyl cylindrical coordinates by
\begin{equation}
\rho = \sqrt{r^2-2mr} \sin \theta \,, \quad
z = (r-m) \cos \theta \,.
\end{equation}
The line element of the magnetic LWP metric~\eqref{LWP-m} in spherical coordinates reads
\begin{equation}
\label{LWP-m-r-theta}
{ds}^2 =  \bar{f} ( d\phi - \bar{\omega} d\tau )^2 + \bar{f}^{-1} \biggl[ -\rho^2 {d\tau}^2 + e^{2\bar{\gamma}} (r^2-2mr+m^2\sin^2\theta) \biggl( \frac{{d r}^2}{r^2-2mr} + {d \theta}^2 \biggr) \biggr] \,.
\end{equation}
Comparing the seed~\eqref{schw} to the above metric we can identify the seed structure functions
\begin{equation}
\bar{f}_0(r,\theta) = r^2 \sin^2 \theta \,, \quad
\bar{\omega}_0(r,\theta) =0 \,.
\end{equation}
The value of $\bar{\gamma}_0$ is not fundamental because it is invariant under Ehlers transformations, however we make explicit it for completeness
\begin{equation}
\label{gamma0}
\bar{\gamma}_0(r,\theta) = \frac{1}{2} \log \biggl(\frac{r^4 \sin^2 \theta}{r^2-2mr+m^2\sin^2\theta} \biggr) \,.
\end{equation}
From definition~\eqref{h}, it is clear that $\bar{h}$ is at most constant, but that constant can be reabsorbed in a coordinate transformation.
Therefore without loss of generality we can choose $\bar{h}_0(r,\theta)=0$.
Finally the seed Ernst gravitational potential takes the form
\begin{equation}
\bar{\Er}_0(r,\theta) = \bar{f}_0(r,\theta) \,.
\end{equation}
The new solution, expressed in terms of the complex potential, is generated via the Ehlers transformation $E$~\eqref{Ehlers-e}, which gives
\begin{equation}
\label{sol-Er}
\bar{\Er}(r,\theta) = \frac{\bar{\Er}_0}{ 1 + i \jmath \bar{\Er}_0} = \frac{r^2 \sin^2 \theta}{1+ i \jmath\, r^2 \sin^2 \theta} \,.
\end{equation}
Note that in case we had used the LWP metric defined in~\eqref{LWP-e}, we would have obtained, via the Ehlers transformation acting on the Schwarzschild seed, the Taub--NUT spacetime, as explained in~\cite{ehlers-marcoa}\footnote{Possibly the solution~\eqref{sol-Er} can be obtained from the \emph{electric} version of the LWP metric~\eqref{LWP-e} by using another variant of the Ehlers transformation $\bar{E}$ composed by the conjugation $W$ and the Ehlers transforamtion $E$ as follows:
$\bar{E} = W \circ E \circ W$.}.

The solution, in metric form, is extracted from the definition of the transformed Ernst potential $\bar{\Er}$.
Hence, according to $\bar{\Er} = \bar{f} + i \bar{h}$, we find
\begin{equation}
\bar{f}(r,\theta) = \frac{r^2 \sin^2 \theta}{1+\jmath^2r^4\sin^4\theta} \,,
\quad \bar{h}(r,\theta) = \frac{\jmath \, r^4 \sin^4 \theta}{1+\jmath^2r^4\sin^4\theta} \,.
\end{equation}
$\bar{\omega}$ has to be found from the definition of $\bar{h}$, as in~\eqref{h}.
At this scope, for sake of completeness, we write down the relevant differential operators in the spherical coordinates
\begin{align}
\label{gradient}
\vec{\nabla} f(r,\theta) & = \frac{1}{\sqrt{r^2-2mr + m^2\sin^2\theta}}
\biggl[ \vec{e}_r \sqrt{r^2-2mr} \frac{\p f(r,\theta)}{\p r}  + \vec{e}_\theta \frac{\p f(r,\theta)}{\p \theta} \biggr]  \,, \\ \label{laplacian}
\nabla^2 f(r,\theta) & = \frac{1}{\sqrt{r^2-2mr + m^2\sin^2\theta}}
\biggl\{ \frac{\p}{\p r} \biggl[ (r^2-2mr) \frac{\p f(r,\theta) }{\p r} \biggr] + \frac{\p}{\p \theta} \biggl[ \frac{\p f(r,\theta) }{\p \theta} \biggr] \biggr\} \,.
\end{align}
The result for the metric function is
\begin{equation}
\bar{\omega}(r,\theta) = 4 \jmath (r- 2m) \cos\theta + \omega_0 \,,
\end{equation}
where $\omega_0$ is an integration constant related to the frame of reference.
Thus, recalling that $\bar{\gamma}$ is not affected by the Ehlers map, the full new metric is
\begin{equation}
\label{bh-rot-universe}
{ds}^2 = F(r,\theta)
\biggl[ - \biggl( 1 - \frac{2m}{r}\biggr) {dt}^2
+ \frac{dr^2}{1 - \frac{2m}{r}} + r^2 d\theta^2 \biggr] \\
+ \frac{r^2 \sin^2 \theta}{F(r,\theta)}
\biggl\{ d\varphi + \big[4\jmath(r-2m)\cos \theta + \omega_0 \big] \ dt \biggr\}^2 \,,
\end{equation}
where we have defined the function
\begin{equation}
F(r,\theta) \coloneqq 1+\jmath^2r^4 \sin^4 \theta \,,
\end{equation}
and renamed $\tau=t$, $\phi=\varphi$.

We can immediately observe that the new metric~\eqref{bh-rot-universe} represents a non-asymptotically flat deformation of the Schwarzschild black hole.
Its structure is quite similar to the Schwarzschild--Melvin spacetime \cite{ernst-magnetic}, and indeed the magnetic Ehlers map that we have used works in a similar fashion as the Harrison transformation.
For this reason we do not expect that the new parameter can be considered either as an hair or a conserved charge of the black hole.  
The physical description of~\eqref{bh-rot-universe} will be analysed in detail in the next section.

Starting with a more general seed we can obtain generalisations of the metric built above.
In Sec.~\ref{sec:kerr} we embed the Kerr black hole in the swirling background, while in Appendix~\ref{app:zipoy} we generate the Zipoy--Voorhees extension of the spacetime~\eqref{bh-rot-universe}. 

\section{Schwarzschild black hole in a swirling universe}
\label{sec:analysis}

In the interpretation of the new spacetime~\eqref{bh-rot-universe} a fundamental point comes from the physical meaning of the new parameter $\jmath$, which defines the behaviour of the gravitational background and, at best of our knowledge, is  unknown.
Thus, we firstly analyse the background metric obtained by turning off the mass parameter $m$ in Eq.~\eqref{bh-rot-universe} and then, in section~\ref{sec:analysis-blackhole}, we study the full black hole solution in its surrounding universe.

\subsection{Analysis of the background spacetime}
\label{sec:analysis-background}

When the mass parameter $m$ vanishes the black hole disappears and we are left with the rotating gravitational background only
\begin{equation}
\label{background-metric}
{ds}^2 = F
\bigl( - {dt}^2 + {dr}^2 + r^2 {d\theta}^2 \bigr) + F^{-1} r^2 \sin^2 \theta
( d\varphi + 4\jmath \, r\cos \theta \ dt )^2 \,.
\end{equation}
In cylindrical coordinates
\begin{equation} 
\label{cyly-coord}
\rho = r \sin \theta \,, \quad
z = r \cos \theta \,, 
\end{equation}
the background takes the simpler form
\begin{equation}
\label{background-cyl}
{ds}^2 = \bigl(1+\jmath^2 \rho^4\bigr)\bigl(-{dt}^2 + {d\rho}^2 + {dz}^2\bigr)
+ \frac{\rho^2}{1+\jmath^2\rho^4}(d\varphi + 4\jmath z dt)^2 \,.
\end{equation}
Such a metric has the very same form of the one presented in Appendix C of~\cite{gibbons-ergo}:
however, it was not studied in that reference.

Symmetries of the metric are given by the four independent Killing vectors
\begin{equation}
\p_t \, \qquad
\p_\varphi \,, \qquad
z \p_t +t\p_z - 2 \jmath (t^2+z^2)\p_\varphi \,, \qquad
\p_z-4\jmath t\p_\varphi \,,
\end{equation}
and by the non-trivial Killing--Yano 2-form
\begin{equation}
-4 \jmath \rho z \, dt \wedge d\rho
+ \jmath \rho^2 \bigl(1+\jmath^2 \rho^4\bigr) \, dt \wedge dz
+ \rho \, d\rho \wedge d\varphi \,.
\end{equation}
The spacetime~\eqref{background-cyl} belongs to the Petrov type D class~\cite{stephani}, and all its Newman--Penrose spin coefficients are null.
These features allow us to infer that the metric~\eqref{background-cyl} belongs to the Kundt class (cf. Table 38.9 of~\cite{stephani}).
We can indeed explicitly express the background metric~\eqref{background-cyl} in the standard Kundt form, by performing the rescaling $t\to \jmath t$ and the change of coordinates
\begin{equation}
q = 2\jmath z \,, \quad
p = \rho^2 \,.
\end{equation}
Metric~\eqref{background-cyl} then boils down to (after a rescaling of the conformal factor)
\begin{equation}
\label{kundt}
{ds}^2 = (\gamma^2 + p^2) \bigl( -{dt}^2 + {dq}^2 \bigr)
+ \frac{\gamma^2 + p^2}{\gamma^2 p} {dp}^2
+ \frac{\gamma^2 p}{\gamma^2 + p^2} ( d\varphi + 2\gamma q dt )^2 \,,
\end{equation}
where we have defined $\gamma=1/\jmath$.
This metric is equivalent to (16.27) of~\cite{Griffiths:2009dfa}, once we put
$m=e=g=\Lambda=\alpha=\epsilon_2=k=0$, $\epsilon_0=1$ and $n=\gamma^2/2$.
One can check that the consistency constraints of~\cite{Griffiths:2009dfa} are indeed satisfied.

The metric~\eqref{kundt}, despite being known for a long time (it was probably discovered by Carter in~\cite{Carter:1968ks}), does not have a clear physical interpretation.
In particular, the physical significance of the parameter $\gamma$ (i.e.~$\jmath$) is not known:
it has been called ``anti-NUT'' parameter by Pleba\'nski in~\cite{Plebanski:1975xfb} because of its resemblance with the NUT parameter in the Pleba\'nski--Demia\'nski spacetime~\cite{Plebanski:1976gy}\footnote{Actually, there exists an analogical connection between the rotating parameter $\jmath$ and the NUT parameter $\ell$ that will be exploited in section~\ref{sec:double-wick}.}.
A generalisation in the presence of the cosmological constant and some possible interpretation of this background is done in section \ref{sec:double-wick}. 

The background metric, written in the form~\eqref{kundt}, has a notable limit to a Levi-Civita metric
\begin{equation}
{ds}^2 = - \hat{\rho}^{4\sigma} {d\tau}^2 + \hat{\rho}^{4\sigma(2\sigma-1)} \bigl(d\hat{\rho}^2 + B^2 {dz}^2 \bigr) + C^2 \hat{\rho}^{2(1-2\sigma)} d\varphi^2 \,,
\end{equation}
in the subcase where $\sigma=1/4$, which, after a change of coordinates\footnote{$\hat{\rho}=\kappa^2p^2$, $\tau = t/\kappa$, $C=1/\kappa$, $B=\sqrt{\kappa}$, $\varphi=q$, $z=\varphi$ and $\kappa=2^{-2/3}$.}, reads
\begin{equation}
\label{levi-civita}
{ds}^2 = p^2 \bigl( -{dt}^2 + {dq}^2 \bigr) + p {dp}^2 + \frac{1}{p} {d\varphi}^2 \,.
\end{equation}
In fact, rescaling the coordinates in~\eqref{kundt} as follows 
\begin{equation}
\label{rescaling}
t \to \gamma^{-2/3} t \,, \quad
q \to \gamma^{-2/3} q \,, \quad
\varphi \to \gamma^{-2/3} \varphi \,, \quad
p \to \gamma^{2/3} p \,,
\end{equation}
and taking the limit $\gamma \to 0$, we recover~\eqref{levi-civita}.
Moreover, thanks to the dilations~\eqref{rescaling}, for large radial coordinates $p$, the background metric~\eqref{kundt} asymptotically approaches a rotating variant of the $\sigma=1/4$ Levi-Civita metric~\eqref{levi-civita}, i.e.
\begin{equation}
{ds}^2 = - p^2 {dt}^2 + 4\gamma^{1/3} \frac{q}{p} dt d\varphi + \frac{1}{p} {d\varphi}^2 + p {dp}^2 + p^2 {dq}^2 \,.
\end{equation}
Clearly, when $\gamma \to 0$, the static Levi-Civita metric~\eqref{levi-civita} is retrieved\footnote{Note that the Levi-Civita metric~\eqref{levi-civita} corresponds to the Kasner metric ${ds}^2= -\tau^2 + \tau^{2 p_1} {dx}^2 + \tau^{2p_2}{dy}^2 + \tau^{2p_3} {dz}^2$, with constraints $p_1+p_2+p_3=1$ and $p_1^2+p_2^2+p_3^2=1$ ($p_1=p_2=\frac{2}{3}, p_3=-\frac{1}{3}$), up to the analytic continuation ($\tau \to i p$, $x \to i t$).}. This is another analogy with the Bonnor-Melvin spacetime \cite{bonnor}.

In order to gain some physical perspective can be useful to investigate, in some detail, the properties of the background metric by inspecting its geodesics.

\subsubsection{Geodesics in the background spacetime}

We define the following geodesic Lagrangian from the background metric~\eqref{background-cyl}
\begin{equation}
\label{lag-background}
\mathscr{L} = \bigl(1+\jmath^2 \rho^4\bigr) \bigl(-\dot{t}^2 + \dot{\rho}^2 + \dot{z}^2\bigr)
+ \frac{\rho^2}{1+\jmath^2\rho^4}\bigl(\dot{\varphi} + 4\jmath z \dot{t}\bigr)^2 \,,
\end{equation}
where the dots stand for the derivatives with respect to an affine parameter $s$.
We can define, via the Killing vectors $\xi = \partial_t$ and $\Phi = \partial_\varphi$, the standard conserved quantities
\begin{equation}
\label{conserved}
-E \coloneqq g_{\mu \nu}u^\mu \xi^\nu \,, \quad
L \coloneqq g_{\mu \nu}u^\mu \Phi^\nu \,,
\end{equation}
where $u^\mu$ is the four-momentum of the test particle, $E$ is the energy and $L$ is the angular momentum.
The explicit definitions for the conserved quantities and the resulting Lagrangian can be found in Appendix~\ref{sec:appback}

The equations of motion derived from the Lagrangian are quite involved.
Qualitative result can be obtained from the normalisation of the four-momentum, i.e.~from equation
$u^\mu u_\mu = \chi$,
with $\chi = -1$ for timelike geodesics and $\chi = 0$ for null geodesics.
The resulting normalisation equation is~\eqref{4mom-BKG}.
For large values of $\rho$ and for fixed $z$, it follows from such equation that
\begin{equation}
\dot{\rho}^2 \approx L^2 \rho^{-2} \,,
\end{equation}
which has solution $\rho(s)\propto \sqrt{2L s}$:
this means that $\rho$ is not limited as the proper time grows.

\begin{figure}[h]
\captionsetup[subfigure]{labelformat=empty}
\centering
\hspace{-0.2cm}
\includegraphics[width=7cm]{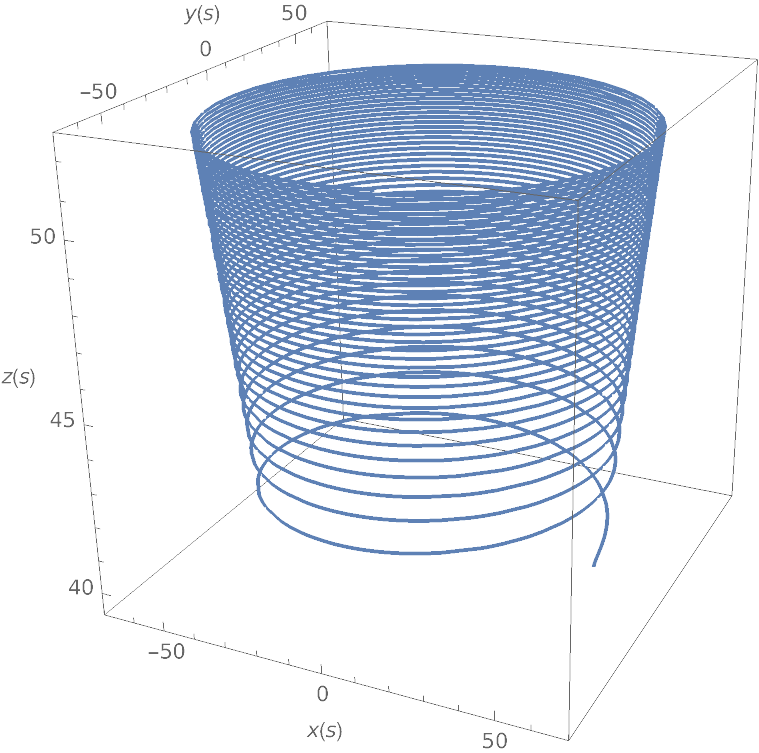}
\caption{\small Geodesic motion of a test particle in the gravitational background ($m=0$) for $ {r(0)=5}, \ \dot{r}(0) = 1, \ z(0) = 40, \ \dot{z}(0) = 1, \ t(0) = 0, \ \dot{t}(0) = 1.4143350824937815,\ \phi(0) = 0,\ \dot{\phi}(0) = 0, \ \jmath=0.1, s_{max} =20, s_{min} =0, \ L=1.81032, \ E=176766$.}%
\label{fig:spiral3d}
\end{figure}

We are interested in analysing the behaviour of $z$ as $\rho$ grows.
We find $\dot{z}^2 = 0$ by letting $\rho\to\infty$ in equation~\eqref{4mom-BKG}, therefore the coordinate $z$ reaches a constant value as $\rho$ approaches infinity.
Moreover, the equation defining $L$, for large values of $\rho$, gives $\phi \approx c^2 L^2 s^2$.
Combining this with the approximate equation for $\rho$, allows one to get the polar equation
$r \approx \sqrt{\frac{2}{c}} \phi^{1/4}$.
Such an equation is the polar form of the generalised Archimedean spiral with exponential 1/4.
Therefore we expect that a geodesic test particle follows a spiral-like path in the $(x,y)$ plane and that the it moves toward a constant value of $z$.
These results are in good agreement with the numerical evaluations, as can be observed from the plot in Fig.~\ref{fig:spiral3d}, which shows the trajectory of a test particle in the $(x,y,z)$ space, where $x=\rho\cos\theta$, $y=\rho\sin\theta$.

The statement that $z$ reaches a constant value can be verified by using the equation of motion for $z$
\begin{equation}
4 \jmath^2 \rho^3\dot{\rho} \dot{z} + (\jmath^2\rho^4+1) \ddot{z}
= \frac{4\rho^2 \jmath \dot{t} (\dot{\varphi} +4 \jmath z \dot{t})}{\jmath^2\rho^4+1} \,.
\end{equation}
By inspecting the equations for $\dot{t}$ and $\dot{\varphi}$ in Appendix~\ref{sec:appback}, one can notice that $\dot{t} \propto 1/\rho^4$ and $\dot{\varphi} \propto \rho^2$ as $\rho\to\infty$, therefore the r.h.s.~of the latter equation can be neglected for large values of $\rho$:
\begin{equation}
4 \jmath^2\rho^3 \dot{\rho} \dot{z} +  (\jmath^2\rho^4+1) \ddot{z} = 0 \,.
\end{equation}
Moreover, $1+\jmath^2 \rho^4 \approx \jmath^2 \rho^4$ as $\rho$ approaches infinity.
By using the approximation $\rho(s) \approx \sqrt{2Ls}$ found above, the equation becomes
\begin{equation}
 2 \dot{z} + s \ddot{z} = 0 \,,
\end{equation}
whose solution is
\begin{equation}
z(s) = \frac{D}{s} + C \,,
\end{equation}
where $C$, $D$ are integration constants.
This result clearly shows that $z$ becomes constant as $s$ approaches infinity.

\subsection{Black hole solution}
\label{sec:analysis-blackhole}

\subsubsection{Physical properties}

The full black hole metric~\eqref{bh-rot-universe}, that we report here for convenience
\begin{equation}
\label{swirling-bh}
{ds}^2 = (1+\jmath^2r^4 \sin^4 \theta)
\biggl[ - \biggl( 1 - \frac{2m}{r}\biggr) {dt}^2
+ \frac{dr^2}{1 - \frac{2m}{r}} + r^2 d\theta^2 \biggr] \\
+ \frac{r^2 \sin^2 \theta}{1+\jmath^2r^4 \sin^4 \theta}
\biggl\{ d\varphi + \big[4\jmath(r-2m)\cos \theta + \omega_0 \big] \ dt \biggr\}^2 ,
\end{equation}
is a two parameters metric, with $m$ and $\jmath$ related to the mass of the black hole and the angular velocity of the background, respectively.

In view of the previous section the spacetime~\eqref{swirling-bh} can be interpreted as a Schwarzschild black hole immersed into a swirling background.
The main causal structure is similar to the Schwarzschild case, as can be readily understood by looking at some $\theta=$ constant slices of the conformal diagram.
For instance, the cases $\theta=\{0,\pm \pi\}$ precisely retrace the static spherically symmetric black hole.

\begin{figure}[h]
\captionsetup[subfigure]{labelformat=empty}
\centering
\hspace{-1cm}
\subfloat[\hspace{1cm} $m=1$, $\jmath=0.05$]{{\includegraphics[scale=0.25]{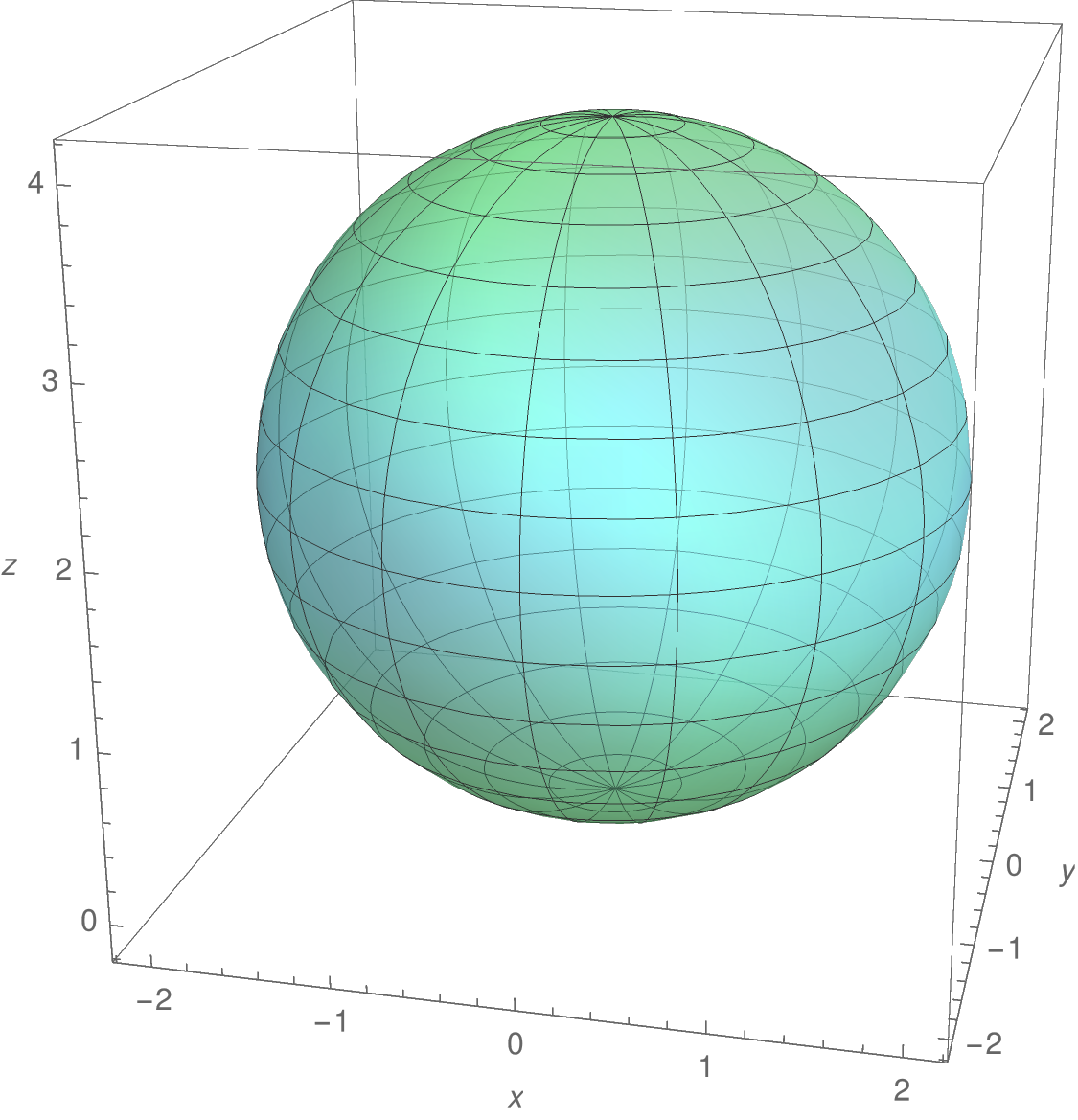}}}%
\subfloat[\hspace{1.5cm} $m=1$, $\jmath=0.1$]{{ \hspace{0.5cm} \includegraphics[scale=0.25]{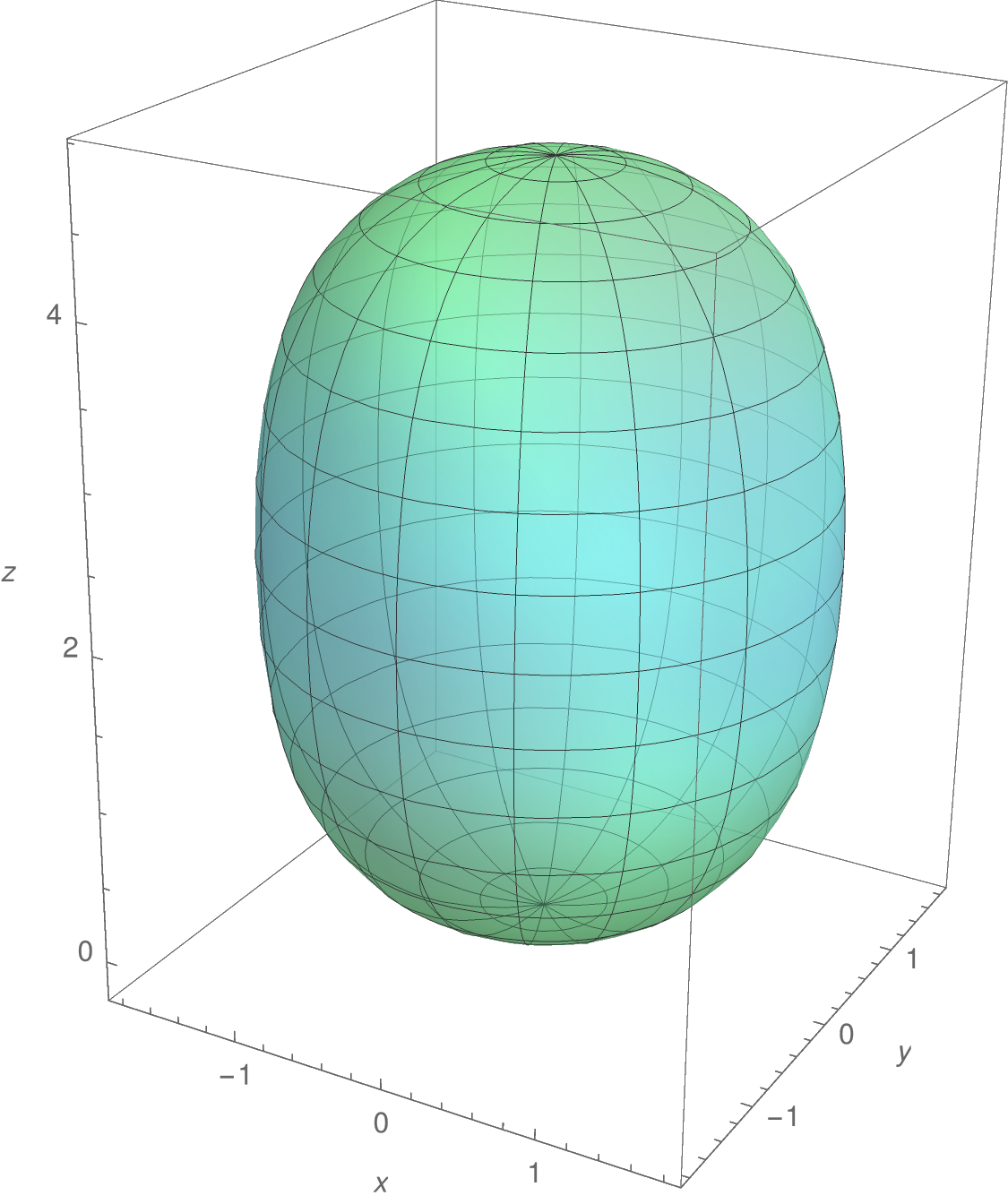}}}%
\subfloat[\hspace{1.5cm} $m=1$, $\jmath=0.3$]{{ 
\hspace{1cm}
\includegraphics[scale=0.25]{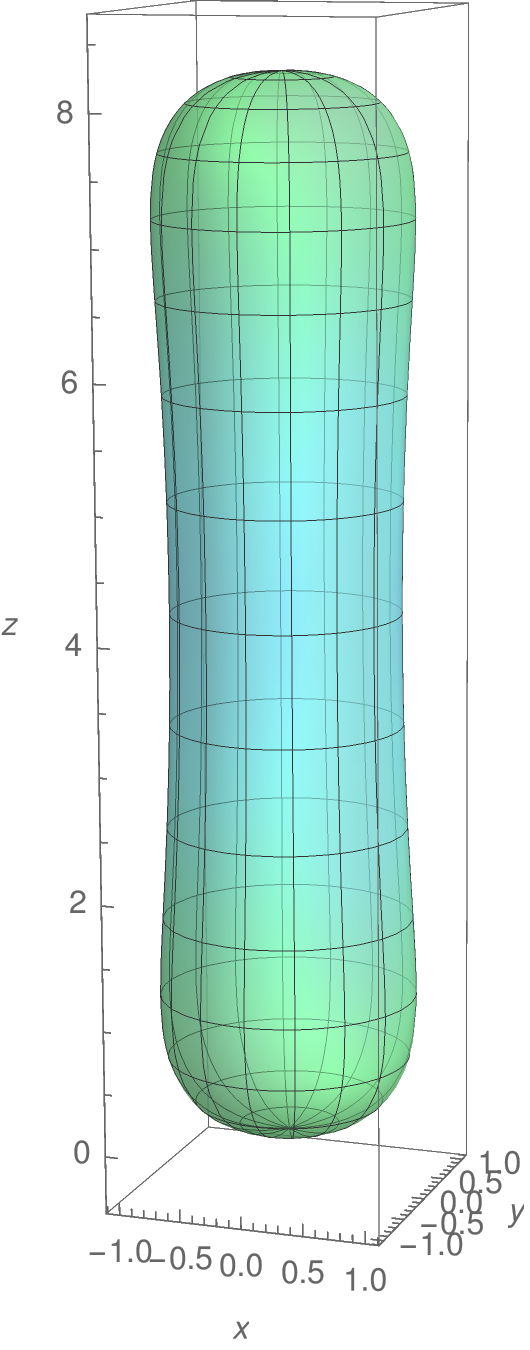}}}
\caption{\small Embedding in Euclidean three-dimensional space $\mathbb{E}^3$ of the event horizon of the black hole distorted by the rotating background, for three different values of the background rotational parameter $\jmath$.}%
\label{picture-horizons}
\end{figure}

Indeed the metric~\eqref{swirling-bh} is characterised by a coordinate singularity located at $r=2m$, which identifies the event horizon of the black hole.
This latter is a Killing horizon that has the same significance of the standard Schwarzschild horizon.
The presence of the rotating background deforms the horizon geometry, making it more oblate, while maintaining exactly the same surface of the Schwarzschild black hole, for the same values of the mass parameter $m$.
In Fig.~\ref{picture-horizons} the deformation is pictured for different intensities of the rotating gravitational background, governed by the new parameter $\jmath$ introduced by the Ehlers transformation.

The solution~\eqref{swirling-bh} is free from axial conical singularities:
to verify this, it is sufficient to consider the ratio between the perimeter of a small circle around the $z$-axis, both for $\theta=0$ and $\theta=\pi$, and its radius.
Such a ratio must be equal to $2\pi$, in case one wants to avoid angular defects.
It turns out that, for the metric (\ref{swirling-bh}), the ratios in the two limits are equal to $2\pi$
\begin{equation}
\lim_{\theta\to0} \frac{1}{\theta}\int_0^{2\pi} \sqrt{\frac{g_{\varphi\varphi}}{g_{\theta\theta}}
}d\varphi
= 2\pi
= \lim_{\theta\to \pi} \ \frac{1}{\pi-\theta}\int_0^{2\pi} \sqrt{\frac{g_{\varphi\varphi}}{g_{\theta\theta}}}d\varphi \,.
\end{equation}
The metric function $\omega(r,\theta)$, as in the electric LWP ansatz (\ref{LWP-e}),  is regular both asymptotically and on the symmetry axis, thus implying the absence of Misner strings or NUT charges.
It is not only continuous, as we can appreciate by the following limits
\begin{equation}
\lim_{\theta\to0} \ \frac{g_{t \varphi}}{g_{tt}}
= \lim_{\theta\to\pi} \ \frac{g_{t\varphi}}{g_{tt}}
= 0 \,,
\end{equation}
but also its first and second derivative are continuous.

A peculiar characteristic of this metric is that the angular velocity $\Omega$ on the $z$-axis is not constant, and it increases in opposite directions in the two hemispheres
\begin{subequations}
\label{angular-velocity}
\begin{align}
\Omega\big|_{\theta = 0} & = \lim_{\theta \rightarrow 0} \biggl( - \frac{g_{t\varphi}}{g_{\varphi\varphi}} \biggr)
= -4\jmath (r - 2m) + \omega_0 \,, \\
\Omega\big|_{\theta = \pi} & = \lim_{\theta \rightarrow \pi} \biggl( -\frac{g_{t\varphi}}{g_{\varphi\varphi}} \biggr)
= 4\jmath (r - 2m) +\omega_0 \,.
\end{align}
\end{subequations}
This is a feature shared with magnetised Reissner--Nordstr\"om and magnetised Kerr black holes solutions~\cite{gibbons-ergo}.

The frame dragging of the whole spacetime is given by~\cite{Poisson:2009pwt}
\begin{equation}
\frac{d\phi}{dt} =
- \frac{g_{t\varphi}}{g_{\varphi\varphi}} = - 4 \jmath ( r -2 m)\cos \theta + \omega_0 \,.
\end{equation}
Hence outside the event horizon, for $r > 2m$, the angular velocity coincides with the asymptotic one $\omega_0$ for $\theta = \frac{\pi}{2}$, while for $\theta \in (\frac{\pi}{2},\pi)$ it is bigger than $\omega_0$ and for $\theta \in (0,\frac{\pi}{2})$ it is smaller than $\omega_0$. 
It is easy to verify that for $r \to \infty$ the angular velocity grows unbounded and that it is equal to $\omega_0$ on the event horizon:
this would lead to the conclusion that there exist superluminal observers, since the value of the gravitational dragging can easily exceed 1 (i.e.~the speed of light, in our units) and, then, it would violate causality.
In this perspective, let us study the possible occurrence of closed timelike curves (CTCs):
considering~\eqref{swirling-bh}, curves in which $t$, $r$ and $\theta$ are constants are characterised by 
\begin{equation}
{ds}_{t,r,\theta=\text{const}}^2 = F^{-1}(r,\theta) \, r^2 \sin^2\theta {d\phi}^2 \,.
\end{equation}
Such intervals are always space-like since the expression is always positive.
Therefore there are no CTCs and there are no related causality issues:
thus the ``paradox'' of the superluminal observers can be justified with the bad choice of the coordinates\footnote{As it happens, for example, for the Alcubierre spacetime~\cite{Alcubierre:1994tu}}.
A set of coordinates which is adapted to timelike observers does not experience an unbounded growth of the angular velocity, as we will see explicitly when studying the geodesics of the spacetime.

The Kretschmann scalar $R_{\mu\nu\rho\sigma}R^{\mu\nu\rho\sigma}$ shows that $r=2m$ is a coordinate singularity, while it is divergent for $r=0$, as in the case of the static spherically symmetric black hole in pure General Relativity.
In particular, as $r\to0$ we find
\begin{equation}
R_{\mu\nu\rho\sigma}R^{\mu\nu\rho\sigma} \approx
\frac{48 m^2}{r^6} \,,
\end{equation}
which is exactly the Kretschmann scalar for the Schwarzschild spacetime.
On the other hand, the scalar invariant decays faster than the Schwarzschild metric for large radial distances, indeed one finds, as $r\to\infty$,
\begin{equation}
R_{\mu\nu\rho\sigma}R^{\mu\nu\rho\sigma} \approx
\frac{192}{\jmath^4 \sin^{12}\theta \, r^{12}} \,,
\end{equation}
therefore the solution~\eqref{swirling-bh} is locally asymptotically flat.
We finally notice that for $\theta=0,\,\pi$ the spacetime has an asymptotic constant curvature:
we find
\begin{equation}
R_{\mu\nu\rho\sigma}R^{\mu\nu\rho\sigma} \big|_{\theta=0,\,\pi} \approx -192 \jmath^2 \quad
\text{ as } r \to \infty \,,
\end{equation}
thus we see that on the $z$-axis, the spacetime is asimptotically of negative constant curvature. The full Kretschmann scalar can be found in appendix \ref{sec:appkre}.

\begin{figure}[h]
\captionsetup[subfigure]{labelformat=empty}
\centering
\hspace{0.2cm}
\subfloat[$m=1$, $\jmath=0.3$]{{\hspace{-0.1cm} \includegraphics[width=8cm]{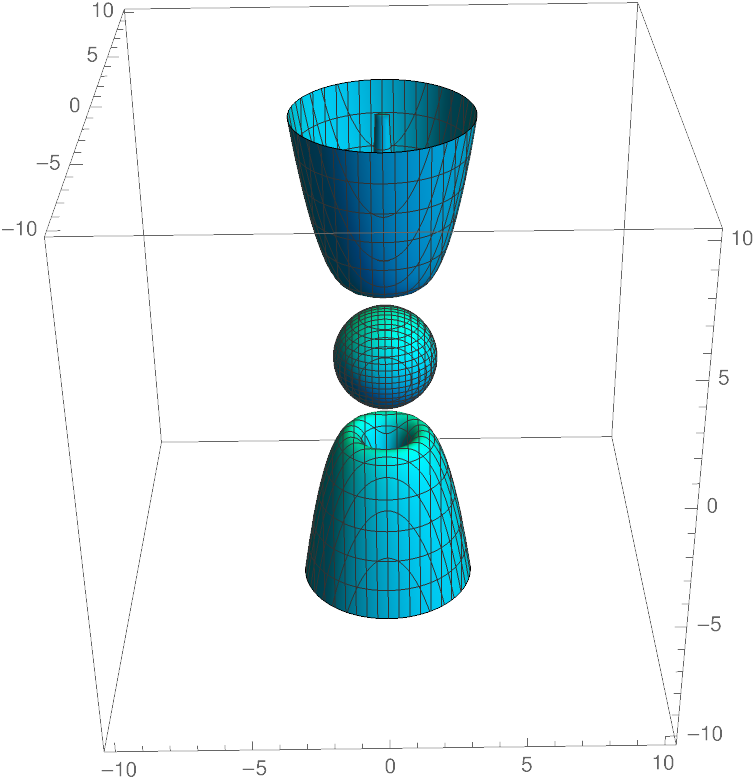}}}
\caption{\small Ergoregions for the black hole embedded in a rotating universe, with parameters $m=1$, $\jmath=0.3$ and $\omega_0=0$.
The ergoregions extend to infinity in the positive and negative $z$ directions, independently of the choice of the parametrisation for the integrating constants.}%
\label{picture-ergoregions3D}
\end{figure}

\paragraph{Ergoregions.}
It is clear, just by inspection, that the $g_{tt}$ component of the metric~\eqref{swirling-bh} becomes null on the event horizon, and that outside the horizon is not everywhere negative. 
Therefore the spacetime presents some ergoregions, analogously to Kerr or magnetised Reissner--Nordstr\"om black holes~\cite{gibbons-ergo}.
To analyse these regions it is convenient to use the cylindrical coordinates as defined in~\eqref{cyly-coord} to expand, for large $z$, the $g_{tt}$ part of the metric as follows
\begin{equation}
g_{tt}(\rho,z) \approx \frac{16 \jmath^2 \rho^2 z (z- 4m)}{1+\jmath^2 \rho^4} \,.
\end{equation}
Hence, the ergoregions are located not only in the proximity of the event horizon, as in Kerr spacetime, but also close to the $z$-axis, for large values of $z$.
A numerical analysis of the function $g_{tt}$ is represented in Fig.~\ref{picture-ergoregions3D}.
It shows how the ergoregions extend to infinity around the polar axis, independently of the values of the integrating constants of the solution $(m,\jmath)$.
This behaviour is similar to what happens for magnetised rotating black holes~\cite{gibbons-ergo}.

\subsubsection{Petrov type}

A standard procedure to determine the Petrov type of a spacetime consists in computing the Weyl tensor in a null tetrad basis.
We define a frame by
\begin{subequations}
\begin{align}
e^0 & = F^{1/2} \biggl(1-\frac{2m}{r}\biggr)^{1/2} \, dt \,, \quad
e^1 = F^{1/2} \biggl(1-\frac{2m}{r}\biggr)^{-1/2} \, dr \,, \quad
e^2 = r F^{1/2} \, d\theta \,, \\
e^3 & = r\sin\theta F^{-1/2} \bigl\{ d\varphi + \big[4\jmath(r-2m)\cos \theta + \omega_0 \big] \ dt \bigr\} \,.
\end{align}
\end{subequations}
Given such a frame, the null tetrad is found as
\begin{equation}
k_\mu = \frac{1}{\sqrt{2}} \bigl(e^0_\mu + e^3_\mu\bigr) \,, \quad
l_\mu = \frac{1}{\sqrt{2}} \bigl(e^0_\mu - e^3_\mu\bigr) \,, \quad
m_\mu = \frac{1}{\sqrt{2}} \bigl(e^1_\mu - i e^2_\mu\bigr) \,, \quad
\bar{m}_\mu = \frac{1}{\sqrt{2}} \bigl(e^1_\mu + i e^2_\mu\bigr) \,.
\end{equation}
It is now possible to compute the components of the Weyl tensor in the null basis, as
\begin{subequations}
\begin{align}
\Psi_0 & = C_{\mu\nu\rho\sigma} k^\mu m^\nu k^\rho m^\sigma \,, \quad
\Psi_1 = C_{\mu\nu\rho\sigma} k^\mu l^\nu k^\rho m^\sigma \,, \quad
\Psi_2 = C_{\mu\nu\rho\sigma} k^\mu m^\nu \bar{m}^\rho l^\sigma \,, \\
\Psi_3 & = C_{\mu\nu\rho\sigma} l^\mu k^\nu l^\rho \bar{m}^\sigma \,, \quad\quad\!
\Psi_4 = C_{\mu\nu\rho\sigma} l^\mu \bar{m}^\nu l^\rho \bar{m}^\sigma \,,
\end{align}
\end{subequations}
where $C_{\mu\nu\rho\sigma}$ is the Weyl tensor.

One can easily show that $\Psi_1=\Psi_3=0$, while the other components are more involved.
The inspection of the scalar invariants
\begin{equation}
I = \Psi_0\Psi_4 - 4 \Psi_1\Psi_3 + 3\Psi_2^2 \,, \quad
J = \det
\begin{pmatrix}
\Psi_0 & \Psi_1 & \Psi_2 \\
\Psi_1 & \Psi_2 & \Psi_3 \\
\Psi_2 & \Psi_3 & \Psi_4
\end{pmatrix}
\,,
\end{equation}
reveals that $I^3\neq 27J^2$:
this implies that the spacetime is algebraically general~\cite{stephani,Griffiths:2009dfa}.
Thus, the spacetime belongs to the general Petrov type I, contrary to its background~\eqref{background-metric} or its generating seed, which are both type D.
Further, this result shows that the new black hole~\eqref{swirling-bh} does not belong to the Pleba\'nski--Demia\'nski class of spacetimes~\cite{Plebanski:1976gy}.

\subsubsection{Geodesics}

We follow the same strategy as in the background case and define, from the metric~\eqref{swirling-bh}, the following Lagrangian (dropping the inessential $\omega_0$ term)
\begin{equation}
\label{lag-bh}
\mathscr{L} =  F
\biggl[ - \biggl( 1 - \frac{2m}{r}\biggr) \dot{t}^2
+ \frac{\dot{r}^2}{1 - \frac{2m}{r}} + r^2 \dot{\theta}^2 \biggr] \\
+ F^{-1} r^2 \sin^2 \theta
\Bigl[ \dot{\varphi} + 4\jmath(r-2m)\cos \theta \, \dot{t} \Bigr]^2 \,,
\end{equation}
Proceeding in the same way as the background metric, we obtain the conserved charges equations and the four-momentum normalisation equations, reported in Appendix~\ref{sec:appbh}.

We can extract some qualitative information, especially regarding the quantity $(r-2m)\cos\theta$ that appears in the gravitational dragging.
For stable orbits $r$ is limited, hence the quantity $(r-2m)\cos\theta$ is limited as well.
For unstable orbits we analyse the geodesic motion as $r$ reaches infinity, thus considering large values of $s$.
We notice that $\dot{t} \approx 0$ and $\dot{\varphi} \approx \jmath^2 L r^2 \sin^2\theta$, as $r\to\infty$, and moreover $1- \frac{2m}{r} \approx 1$ and $F \approx \jmath^2 r^4 \sin^4\theta$.
These approximations simplify the Lagrangian~\eqref{lag-bh}, that takes the form
\begin{equation}
\mathscr{L} \approx \jmath^2 r^4 \sin^4\theta
\bigl(
\dot{r}^2 + r^2 \dot{\theta}^2 \bigr)
+ \jmath^2 L^2 \,.
\end{equation}
The constant term is inessential and can be neglected.
By changing to polar coordinates
$x = r \sin \theta$ $y = r \cos \theta$,
the Lagrangian boils down to
\begin{equation}
\mathscr{L} \approx \jmath^2 x^4 \bigl( \dot{x}^2 + \dot{y}^2 \bigr) \,.
\end{equation}
Being the Lagrangian independent on $y$, we find the conserved quantity
\begin{equation}
A = \jmath^2 x^4 \dot{y}^2 \,.
\end{equation}
This result can be plugged into the Lagrangian, and by noticing that the Lagrangian does not depend explicitly on $s$, $d\mathscr{L}/ds=0$, the following equation is derived:
\begin{equation}
\jmath^2 x^4 \dot{x}^2 + \frac{A^2}{\jmath^2 x^4}
= B \,,
\end{equation}
where $B$ is a real constant.
From the last equation we find
\begin{equation}
\dot{x} = \frac{\sqrt{B \jmath^2 x^4 - A^2}}{\jmath^2 x^4} \approx \frac{\sqrt{B}}{\jmath x^2} \,,
\end{equation}
where the numerator $\sqrt{B \jmath^2 x^4 - A^2}$ depends on $x$ which, by our change of coordinate, is proportional to $r$, so when $r$ approaches infinity so does $x$.
Therefore the constant $A^2$ underneath the square root can be neglected, thus justifying the approximation.
Finally we get, by integrating,
\begin{equation}
x(s) = \biggl(\frac{3 \sqrt{B}}{\jmath} s
+ C \biggr)^{\frac{1}{3}} \,,
\end{equation}
with $C$ real constant.
So $x \to \infty$ as $s \to \infty$, which means that $\dot{y} \approx 0$ and $y \approx$ const.
These results can now be plugged into the formula for the gravitational dragging, which gives
\begin{equation}
- \frac{g_{t\varphi}}{g_{\varphi\varphi}}
= - 4 \jmath \biggl(y - 2m \frac{y}{ \sqrt{x^2 + y^2}}\biggr)
\underset{s\to\infty}{\approx} -4\jmath y \,.
\end{equation}
This result shows that, as $s \to \infty$, the angular velocity approaches a constant value.

\begin{figure}[h]
\captionsetup[subfigure]{labelformat=empty}
\centering
\hspace{-0.2cm}
\subfloat[\centering $m=1$, $\jmath=0$]{{\includegraphics[width=7.5cm]{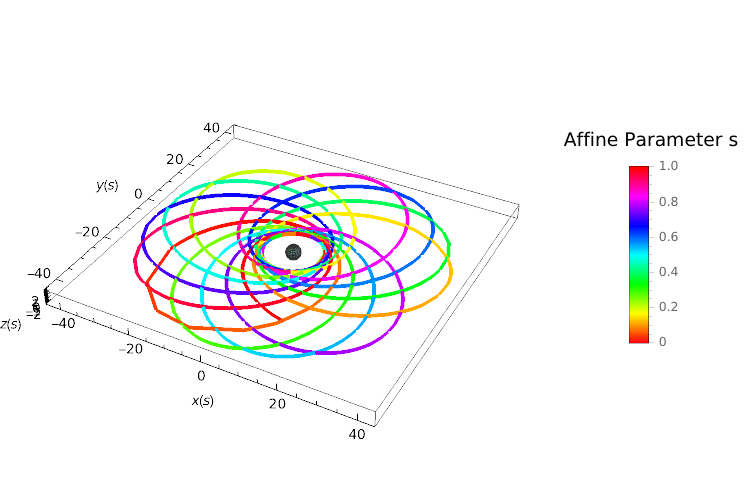}}}%
\subfloat[\centering $m=1$, $\jmath=10^{-8}$]{{\hspace{-0.3cm} \includegraphics[width=7.5cm]{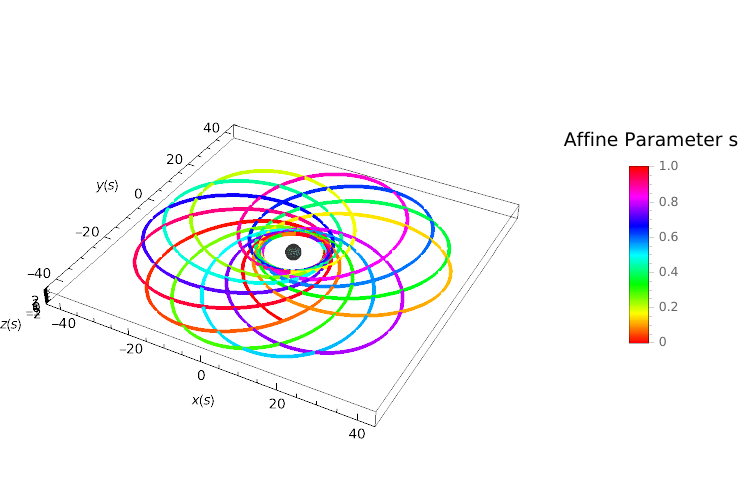}}}
\caption{\small Geodesic motion around the black hole.
The left panel shows the Schwarzschild spacetime, while the right panel shows the new black hole solution~\eqref{swirling-bh}.
The plots share the same initial conditions with $r(0)=8$, $\dot{r}(0) = 0$, $\theta$(0) = $\pi/2$,   $\dot{\theta}(0) = 0 $, $t(0) = 0$, $\dot{t}(0)$ = 1.30703 , $\phi(0) = 0$, $\dot{\phi}(0) = \frac{3\sqrt{2}}{64}$, $s_{max}$ =10000, $s_{min} = 0$, $L = 4.24264$, $E =0.980274$.}
\label{fig:schwVSswirl}
\end{figure}

\begin{figure}
\captionsetup[subfigure]{labelformat=empty}
\centering
\hspace{-0.2cm}
\subfloat[\centering $m=1$, $\jmath=0.01$]{{\includegraphics[width=7.5cm]{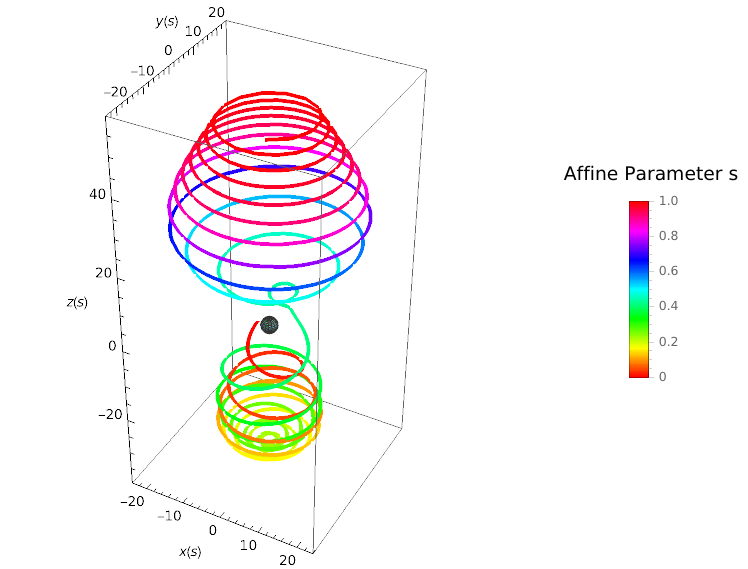}}}%
\subfloat[\centering $m=1$, $\jmath=0.01$]{{\hspace{-0.3cm} \includegraphics[width=7.5cm]{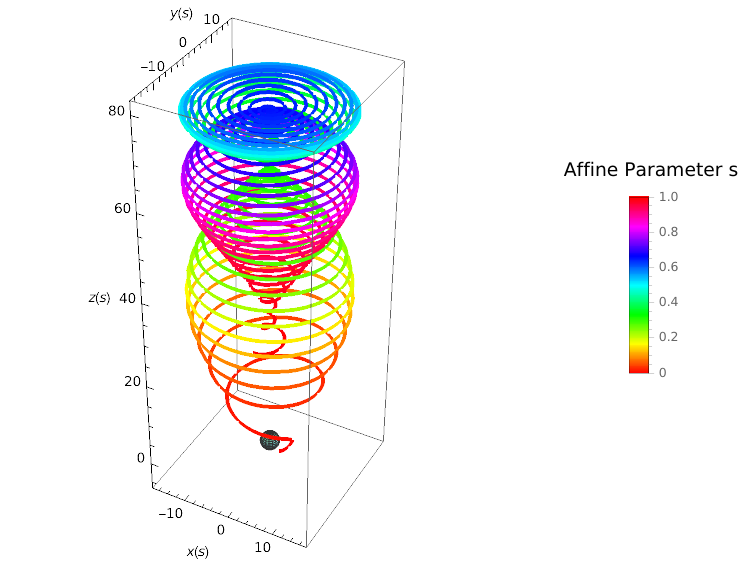}}}
\caption{\small Embedding diagram and geodesics for the new metric~\eqref{swirling-bh} with different initial conditions:  $\theta$(0) = $\pi/2$, $\dot{\theta}(0) = 2 $, $\dot{t}(0)$ = 16.0427 , $\phi(0) = \pi$, $L=2.67831$, $E = 5.39087$ for the l.h.s. diagram and $\theta$(0) = $3\pi/4$, $\dot{\theta}(0) = -4 $, $\dot{t}(0)$ = 24.103 , $\phi(0) = 0$, $L=-1.71434$, $E = 8.0021$ for the r.h.s. diagram.
Both representations share the following data: $r(0)=3$, $\dot{r}(0) = 4$, $t(0) = 0$, $\dot{\phi}(0) = 0.3$, $s_{max} = 145$, $s_{min} =0$.}
\label{fig:bh-geod}
\end{figure}

We also plot the geodesic motion in spacetime~\eqref{swirling-bh}:
this amounts to numerically integrate the geodesic equations reported in Appendix~\ref{sec:appbh}, and the results are shown in Fig.~\ref{fig:schwVSswirl},~\ref{fig:bh-geod} and~\ref{fig:unstable}.
More precisely, Fig.~\ref{fig:schwVSswirl} compares geodesics in Schwarzschild spacetime (i.e.~$\jmath=0$) and geodesics in our swirling spacetime~\eqref{swirling-bh}.
Fig.~\ref{fig:bh-geod} shows the geodesics around the black hole for different initial conditions.
Finally, in Appendix~\ref{sec:appbh}, Fig.~\ref{fig:unstable} pictures unstable geodesic motion for two different values of the test particle angular momentum.

\subsubsection{Charges and thermodynamics}

The total mass of the spacetime can be computed by means of the surface  charges provided by the phase space formalism~\cite{wald,barnich}.
We perturb the metric with respect to the parameters of the solution, and we call that variation $h_{\mu\nu}\coloneqq\delta g_{\mu\nu}$\footnote{In the particular case under consideration the parameter space is spanned by the mass parameter of the black hole $m$ and by the magnitude of the rotational whirlpool dragging, $\jmath$.
Thus the variation takes the form $h_{\mu\nu}= \delta g_{\mu\nu}(m,\jmath) = \partial_m g(m,\jmath) \delta m + \partial_\jmath g(m,\jmath) \delta \jmath$.}.
Then we find the local variation of the charge $K_\xi$ computed along a given Killing direction $\xi^\mu$.

The local variation of the charge must be integrated between the parametric reference background $\bar{\Psi}$ and the actual parametric configuration labelled by $\Psi$, on a $D-2$ dimensional surface $\mathcal{S}$ containing the event horizon $\bigl(d^{D-2}x \bigr)_{\mu\nu} = \frac{1}{2(D-2)!} \epsilon_{\mu\nu\alpha_1...\alpha_{D-2}} dx^{\alpha_1} \wedge \cdots \wedge dx^{\alpha_{D-2}}$.
When the variation of the charge is integrable, all the parametric paths between the reference background and the solution are equivalent\footnote{In case the variation of the charge is not integrable, we still have some gauge degree of freedom in defining the frame of reference, or the normalisation of the time coordinate, to recover integrability.}.

The result gives the total surface charge $Q_\xi$, defined, as in~\cite{barnich,compere}, by
\begin{equation}
Q_{\xi} = \frac{1}{8\pi} \int_{\bar{\Psi}}^{\Psi} \int_\mathcal{S} K_\xi = \frac{1}{8\pi} \int_{\bar{\Psi}}^{\Psi} \int_\mathcal{S} K_\xi^{\mu\nu} \bigl(d^{D-2}x \bigr)_{\mu\nu} \,,
\end{equation}
where
\begin{equation}
K_\xi^{\mu\nu} = \xi^\mu \nabla_\sigma h^{\nu\sigma} - \xi^{\mu} \nabla^\nu h - \xi_\sigma \nabla^\mu h^{\nu\sigma} - \frac{1}{2} h \nabla^\mu \xi^\nu + \frac{1}{2} h^{\sigma \mu} (\nabla_\sigma \xi^\nu - \nabla^\nu \xi_\sigma) \,,
\end{equation}
and where $h\coloneqq h^{\mu}_{\;\;\mu}$.
If we want to compute the mass of the black hole, we have to consider the timelike Killing vector $\xi=\partial_t$, then we find
\begin{equation}
M = Q_{\partial_t} = m \,,
\end{equation}
as for the Schwarzschild black hole.
In this case the presence of the background does not modify the seed black hole mass, similarly to what happens within the context of black holes embedded in an external electromagnetic field~\cite{compere-mass}.
Following this analogy we expect to observe some stronger coupling with the background in case of more general black hole seeds, as it happens in the next section.  

The angular momentum can be found analogously, just considering the Killing vector which generates the rotational symmetry $\partial_\varphi$.
In this case one gets null angular momentum 
\begin{equation}
J=Q_{\partial_\varphi} = 0 \,,
\end{equation}
even though the solution is clearly rotating.
In fact the angular momentum refers just to the dipole term in the rotational multipolar expansion of the metric at large distances.
The fact that the metric is rotating, as its non-diagonal form suggests, can be appreciated by the subsequent terms of the multipolar expansion:
the quadrupole, the octupole, etc\footnote{However, note that in contexts where the asymptotia is not of globally constant curvature, the notion of the gravitational multipolar expansion needs some further analysis to be clearly defined.}.

We compute the entropy and the temperature of the event horizon, in order to study the Smarr law and the thermodynamics of the black hole.
The area of the even horizon is found by integrating the $(\theta,\phi)$ part of the metric, hence
\begin{equation}
\mathcal{A} = \int_0^{2\pi} d\varphi \int_0^\theta d\theta
\sqrt{g_{\theta\theta} g_{\phi\phi}} \big|_{r=2m}
= 16 \pi m^2 \,.
\end{equation}
The entropy is then taken by the Bekenstein--Hawking formula $S = \mathcal{A}/4$.
The validity of the area law also for this unconventional background is confirmed by the conformal field theory dual to the near-horizon geometry of the black hole \cite{marcoa-removal}.
The temperature can be easily obtained via the surface gravity, $\kappa = \sqrt{- (\nabla_\mu\xi_\nu)^2/2}\big|_{r=2m}$, where $\xi=\partial_t$.
We find
\begin{equation}
T = \frac{\kappa}{2\pi} =\frac{1}{8\pi m} \,.
\end{equation}
Note that also the entropy and the temperature of the black hole embedded into the swirling background are unaffected by the spacetime rotation:
they remain the same of the Schwarzschild seed.
Even though the parameter $\jmath$ introduced by the Ehlers transformation can be considered an integration constant for the solution, it is not associated to a new conserved charge, so reasonably it is not present as an additional term in the first law.
Again, this is a peculiarity of the Lie point symmetries we used to generate the solution, a general feature shared with the Harrison transformation, thus with the Schwarzschild--Melvin spacetime~\cite{gibbons-ergo,compere-mass}\footnote{Note that this is true only when the seed does not couple with the background brought in by the transformation, but it does not hold in case of more general seeds.
For more general setting including relaxed boundary conditions and thermodynamic ensembles, additional terms in the first law might appear.}.
Hence, we can easily verify the validity of the Smarr law
\begin{equation}
M = 2 T S \,.
\end{equation}
Further, the conserved charges satisfy the first law of thermodynamics
\begin{equation}
\delta M = T \delta S \,.
\end{equation}

\section{Kerr black hole in a swirling universe}
\label{sec:kerr}

The generating techniques discussed in Sec.~\ref{sec:generation} can be also exploited to embed a rotating black hole in a background endowed with its own rotation.
By using the Kerr metric in Boyer--Lindquist coordinates as a seed, we obtain\footnote{A Mathematica notebook with this solution is added in the arXiv files, for the reader convenience.}
\begin{equation}
{ds}^2 = F (d\varphi - \omega dt)^2 + F^{-1}\biggl[-\rho^2 {dt}^2 + \Sigma\sin^2\theta \biggl(\frac{{dr}^2}{\Delta} + {d\theta}^2\biggr)\biggr] \,,
\end{equation}
where the functions $F^{-1}$ and $\omega$ can be expanded in a finite power series of $\jmath$
\begin{align}
F^{-1}  & = \chi_{(0)} + \jmath\chi_{(1)} + \jmath^2\chi_{(2)} \,, \qquad
\omega = \omega_{(0)} + \jmath\omega_{(1)} + \jmath^2\omega_{(2)} \,,
\end{align}
with 
\begin{subequations}
\begin{align}
\chi_{(0)} & = \frac{R^2}{\Sigma \sin^2\theta} \,, \\
\chi_{(1)} & = \frac{4 a m \, \Xi \cos\theta }{\Sigma \sin^2\theta} \,, \\
\chi_{(2)} & = \frac{4 a^2 m^2 \Xi^2 \cos^2\theta + \Sigma^2 \sin^4\theta}{R^2 \Sigma \sin^2\theta} \,,
\end{align}
\end{subequations}
and 
\begin{subequations}
\begin{align}
\omega_{(0)} & = \frac{2 a m r}{-\Sigma} +\omega_0\,, \\
\omega_{(1)} & = \frac{4 \cos\theta [-a \Omega (r-m) + ma^4 - r^4(r-2m) - \Delta a^2 r]}{-\Sigma } \,, \\
\omega_{(2)} & =
\frac{2m \{3ar^5 - a^5(r+2m) + 2a^3r^2(r+3m) - r^3(\cos^2\theta-6)\Omega + a^2[\cos^2\theta(3r-2m) - 6(r-m)] \Omega\}}{-\Sigma } \,,
\end{align}
\end{subequations}
where
\begin{subequations}
\begin{align}
\Delta & = r^2 - 2mr + a^2 \,, \qquad\qquad\qquad\qquad\qquad\!\!\!\!\!\!\!\!
\rho^2 = \Delta \sin^2\theta \,, \\
\Sigma & = (r^2 + a^2)^2 - \Delta a^2 \sin^2\theta \,, \qquad\qquad\qquad\!\!\!\!\!\!
\Omega = \Delta a \cos^2\theta \,, \\
\Xi & = r^2(\cos^2\theta - 3) - a^2 (1+\cos^2\theta) \,, \qquad
R^2 = r^2 + a^2 \cos^2\theta\,.
\end{align}
\end{subequations}
When $\jmath=0$ we recover the seed metric, i.e.~the Kerr black hole.
For $\jmath\ne0$ we have the direct generalisation of the metric~\eqref{bh-rot-universe}.

However we notice that in this case, because of the spin-spin interaction between the black hole and the background frame dragging, an extra force acts on the axis of symmetry.
But since it is not symmetric on the two hemispheres, the metric is affected by non-removable conical singularities, indeed
\begin{equation}
\lim_{\theta\to0} \frac{1}{\theta}\int_0^{2\pi} \sqrt{\frac{g_{\varphi\varphi}}{g_{\theta\theta}}
}d\varphi
= \frac{2\pi}{(1-4a m \jmath)^2} \neq \lim_{\theta\to \pi} \ \frac{1}{\pi-\theta}\int_0^{2\pi} \sqrt{\frac{g_{\varphi\varphi}}{g_{\theta\theta}}}d\varphi = \frac{2\pi}{(1+4a m \jmath)^2} \,.
\end{equation}
In fact, even though the background spinning parameter $\jmath$ couples to the Kerr angular momentum (for unit of mass) $a$, it is not possible to find a relation among the physical parameters to remove simultaneously both angular defects, unless of course for known subcases such as Kerr, for $\jmath=0$, the spacetime discussed in Sec.~\ref{sec:analysis} for $a=0$, or the rotating background for $m=0$.
The presence of a non-removable conical singularity implies that a cosmic string or a strut (with their $\delta$-like stress-energy-momentum tensor on a portion of the $z$-axis) have to be postulated in order to compensate the ``force'' effect induced by the spin-spin interaction of the black hole with the background, which would tent to add acceleration to the black hole\footnote{The metric considered in this section does not posses the acceleration parameter:
one should work with the rotating C-metric to consistently include the acceleration.
That is why, in this section, the role of the string uniquely results in the effect of compensating the spin coupling.}.

In the case one wants to immerse the Kerr--Newman black hole into this spinning universe, one has to use the charged generalised version of the Ehlers transformation, as described in~\cite{ehlers-marcoa}.

\section{Double-Wick rotation of the background: flat Taub--NUT spacetime}
\label{sec:double-wick}

Given the analogies between the rotating background~\eqref{background-cyl} and the Melvin spacetime, and given that the analytical continuation of the Melvin universe corresponds to the Reissner--Nordstr\"om metric with a flat base manifold, it is natural to inquire about an analog analytical continuation for the rotating background.
At this scope, we implement a double Wick rotation between time and the azimuthal angle $t \to i \phi$, $\varphi \to i \tau $ of the metric~\eqref{background-cyl}.
Redefining the integration constant of the rotating background as $\jmath=m/2\ell^3$, changing the coordinate $\rho = \ell \sqrt{2r/m}$ and after the rescaling of the other three coordinates we obtain
\begin{equation} \label{T-NUT}
{ds}^2 = - \frac{2mr}{r^2+\ell^2} ( dt - 2 \ell \theta d\phi)^2 + \frac{r^2+\ell^2}{2mr} {dr}^2 + (r^2+\ell^2) ({d\theta}^2 + {d\phi}^2 ) \,.
\end{equation}
It is not hard to recognise the Taub--NUT spacetime with a flat, or possibly cylindrical if we keep the azimuthal angle identification, base manifold.
In fact, the flat Taub--NUT metric can be generated via the Ehlers transformation\footnote{In this section we are referring to transformations applied to the $magnetic$ LWP metric, as explained in Sec.~\ref{sec:generation}.}, from the Schwarzschild metric, previously composed with a double-Wick rotation\footnote{While this is true for a generic sign of the constant curvature of the seed base manifold, only the metric with positive curvature can be interpreted as a black hole in Einstein gravity.}.
Note that the Ehlers transformation can be used to build, from the Minkowski seed, the rotating background, just considering $m=0$ in the procedure of Sec.~\ref{sec:generation}.
This is analogous to what happens to the Melvin universe, which can be obtained from Minkowski spacetime through an Harrison transformation and whose double-Wick dual corresponds to the flat Reissner--Nordstr\"om metric.
This fact strengthens the link between the Melvin universe and our rotating background.
Actually this correspondence can be summarized by the following proportion:
\begin{equation}
\label{proportion}
\boxed{\text{Melvin Universe : Harrison transformation = Rotating Universe : Ehlers transformation}}
\end{equation}
This formal analogy can be exploited to build new solutions, even outside the range of the generating technique based on the Lie point symmetries of the Ernst equations.
That is because even if the symmetry transformations such as the Ehlers and the Harrison maps break in the presence of the cosmological constant. However, as noted in~\cite{marcoa-charging}, the Melvin universe can still be generalised when the cosmological constant is not zero, and it still preserves its relation with the flat Reissner--Nordstr\"om metric with a constant curvature base manifold\footnote{Metrics without a topological spherical base manifold are interpreted as black holes in the presence of the cosmological constant.}.
Therefore, thanks to the analogy with the Melvin case, we have in our hand a procedure to generalise the rotating background~\eqref{background-cyl} in the presence of the cosmological constant.
It is sufficient to operate a double-Wick rotation of the cosmological version of the flat Taub--NUT metric~\eqref{T-NUT}
\begin{equation}
{ds}^2 = - \frac{\frac{\Lambda}{3} r^4 + 2\ell^2\Lambda r^2 + 2mr - \ell^4 \Lambda}{r^2+\ell^2} ( dt - 2 \ell \theta d\phi)^2 + \frac{(r^2+\ell^2) {dr}^2}{\frac{\Lambda}{3} r^4 + 2\ell^2\Lambda r^2 + 2mr - \ell^4 \Lambda} + (r^2+\ell^2) ({d\theta}^2 + {d\phi}^2) \,.
\end{equation}
Thus, using the same change of coordinates and parametrisation of the case above, we get 
\begin{equation}
\label{kundt-lambda1}
{ds}^2 =  (1+\jmath^2 \rho^4) \biggl(-{d\tau}^2 + \frac{\rho^2 d\rho^2}{\frac{\Lambda}{4\jmath^2} + \rho^2-\frac{\Lambda}{2}\rho^4 -\frac{\jmath^2 \Lambda}{12}\rho^8} + {dz}^2 \biggr) + \biggl(\frac{\Lambda}{4\jmath^2} + \rho^2-\frac{\Lambda}{2}\rho^4 -\frac{\jmath^2 \Lambda}{12}\rho^8\biggr) \frac{(d\psi + 4\jmath z d\tau)^2}{1+\jmath^2 \rho^4} \,.
\end{equation}
It is not difficult to realise that this metric still corresponds, up to a change of coordinates, to the non-expanding and non-accelerating Kuntdt class of the Pleba\'nski--Demia\'nski family presented in Eq.~(16.26) of~\cite{Griffiths:2009dfa}.
The explicit change of coordinates works as in Sec.~\ref{sec:analysis-background}, namely
$q=2\jmath z$ and $p^2=\rho$, together with the rescaling
$t\to\jmath t$ and the redefinitions $\gamma=1/\jmath$, $\Tilde{\Lambda}=4\Lambda$.
Then, metric~\eqref{kundt-lambda1} becomes
\begin{equation}
\label{kundt-lambda2}
{ds}^2 = (\gamma^2+p^2) \biggl(-{d\tau}^2 + \frac{dp^2}{\mathcal{P}} + {dq}^2 \biggr)
+ \frac{\mathcal{P}}{\gamma^2+p^2} (d\psi + 2\gamma q d\tau)^2 \,,
\end{equation}
where
\begin{equation}
\mathcal{P} = \gamma^4\Tilde{\Lambda} + \gamma^2 p - 2\gamma^2\Tilde{\Lambda} p^2 - \frac{\Tilde{\Lambda}}{3} p^4 \,.
\end{equation}
The latter corresponds to Eq.~(16.26) of~\cite{Griffiths:2009dfa}, where
$m=e=g=\alpha=\epsilon_2=0$, $\epsilon_0=1$, $k=\gamma^4\Tilde{\Lambda}$, $\epsilon=2\gamma^2\Tilde{\Lambda}$ and $n=\gamma^2/2$.

The analogy between the rotating background and the Taub--NUT spacetime can be pushed further:
it is known~\cite{Emparan:2001gm} that the Melvin spacetime corresponds to a couple of magnetically charged Reissner--Nordstr\"om black holes moved toward infinity.
In this sense, the magnetic field which permeates the Melvin spacetime is nothing but the field generated by two black hole sources at infinity.
Thus, it is natural to ask ourselves if a similar construction also holds for the rotating background~\eqref{background-metric}, i.e.~if it can be obtained as a limit of a double black hole metric.

By relying on the above considerations and, more specifically, on the proportion~\eqref{proportion}, the natural candidate for a ``ancestor'' metric is the double-Taub--NUT spacetime with opposite NUT parameters~\cite{KramerNeug}:
the rotation of the two counter-rotating Taub--NUT black holes, once they are pushed at infinity, should produce the rotation of the background that is experienced in the background spacetime.
This interpretation is also consistent with the behaviour of the angular velocity~\eqref{angular-velocity}:
we noticed that the angular velocity increases in opposite directions in the two hemisphere, coherently with the fact that the two black holes rotate in different directions.
Moreover this picture is enforced by the geometry of the ergoregions, since the latter thrive for large values of $z$ on the axis of symmetry. 

\section{Summary and Conclusions}

Being inspired by the work of Ernst that used the Harrison transformation to immerse any stationary and axisymmetric spacetime into the Melvin electromagnetic universe, we have considered, in this article, a symmetry transformation of the Ernst equations that embeds any given seed spacetime into a rotating background.
This transformation, which consists in a proper composition of the electric Ehlers transformation with a couple of discrete symmetries, known as conjugations, allows us to take advantage of the Ernst solution generating technique to non-linearly superpose the Schwarzschild black hole and a swirling universe.
The background geometry can be interpreted as a gravitational whirlpool generated by a couple of counter-rotating sources at infinity.
Its frame dragging transforms the static Schwarzschild metric into a stationary one, removing the asymptotic flatness, but without drastically altering the black hole causal structure nor introducing pathological features.
The analogies between the swirling background and the Melvin universe are numerous, like the metric structure, the ergoregions and the deformations engraved on the event horizon:
in fact the former universe can be considered as the rotating counterpart of the latter.

For this reason we expect that this spinning background can be used as a regularising instrument for metric with conical singularities, exactly as the electromagnetic background brought by the Harrison transformation does.
In the former case to have non-trivial physical effects one needs to exploit the interplay between the coupling of the gravitational field of the seed with the one of the background, as suggested by the analysis of the transformed Kerr metric, in section~\ref{sec:kerr}.
Indeed the interaction between the Kerr parameter $a$ and the background parameter $\jmath$ generates an additional ``force'' that impels the system to accelerate.
Unfortunately the geometry of the spacetime is not general enough to accommodate this physical feature into that metric, yielding a conical singularity which compensates the mutual rotational coupling.
On the other hand we count that the spin-spin interaction between the seed and the background environment can play a relevant role into the regularisation gravitational models which otherwise would be mathematically defective and physically incomplete.
For instance, as Ernst showed that the electromagnetic background can eliminate the conical defect of the accelerating and charged black hole, we foresee that the procedure presented in this paper can remove the axial singularities of the rotating C-metric, providing at the same time a reasonable physical explication for its acceleration, works in this direction are in progress \cite{marcoa-removal}.
Also this model furnishes alternative scenarios for black hole nucleation and pair creation, without relying on the electromagnetic field, as discussed in the literature so far~\cite{strominger,hawking,marcoa-pair}.

Clearly this procedure may be relevant for other systems, not necessarily accelerating, such as balancing multi-black hole sources to reach an equilibrium configuration.
Also in that case the frame dragging of the background can play a role in removing cosmic strings or strut from the singular spacetime.
On the other hand, preliminary studies suggest that the spin-spin interaction between the swirling universe and a Taub--NUT spacetime are not sufficient to mend also the singular behaviour of that metric, i.e.~to remove the Misner string\footnote{Obviously we are referring to the non-compact time representation of the Taub--NUT metric, because when one considers proper periodic identification of the temporal coordinate the spacetime can be regularised.
Unfortunately this latter representation violates causality because of the appearance of closed timelike curves, which makes this picture nonphysical.}.

From a phenomenological point of view our rotating background might be of some interest in the description of black holes surrounded by interacting matter, which produces intense frame dragging, such as the one caused by the collision of counter-rotating galaxies. 

Since this construction is based on a symmetry transformation of the Ernst equations, it can be directly generalised to the Einstein--Maxwell case, to the minimally and conformally coupled scalar field case and, more generally, to scalar-tensor theories such as Brans--Dicke, just by using the adequate Ehlers transformation as described in~\cite{ehlers-marcoa} and~\cite{marcoa-embedding} respectively.
The embedding method presented here may reveal also useful in establishing and improving traversability of wormhole spacetimes.\\

\section*{Aknowledgments}

This article is based on the results of the bachelor thesis~\cite{tesi-riccardo} by R.M. We would like to thank Rogério Capobianco for pointing us some inconsistency in the pictures of the geodesic orbits and for discussion about that issue; for more details about the geodesic motion in swirling background see \cite{Rogerio}. 
This work was partly supported by MIUR-PRIN contract 2017CC72MK003 and by INFN.\\

\appendix

\section{Geodesics}

We report here the explicit expression for the geodesic equation, both for the background and the full black hole metric.

\subsection{Background geodesics}
\label{sec:appback}

The explicit expressions for the definitions of the conserved quantities~\eqref{conserved} are
\begin{equation}
\dot{t} = \frac{E + 4\jmath L z}{1+\jmath^2 \rho^4} \,, \qquad
\dot{\phi} = \jmath^2 L \rho^2 + \frac{L}{\rho^2}
- 4 \jmath z \frac{E + 4\jmath L z}{1 + \jmath^2 \rho^4} \,.
\end{equation}
By substituting these relations into the Lagrangian~\eqref{lag-background}, we get
\begin{equation}
\mathscr{L} =
(1 + \jmath^2 \rho^2)
\biggl[ \frac{L^2}{\rho^2} - \biggl(\frac{E + 4\jmath L z}{1+\jmath^2 \rho^4}\biggr)^2 + \dot{\rho}^2 + \dot{z}^2 \biggr] \,.
\end{equation}
The equation coming from the normalisation of the four-momentum $u_\mu u^\mu = \chi$ is
\begin{equation}
\label{4mom-BKG}
\begin{split}
\dot{\rho}^2 + \dot{z}^2 & =
\frac{1}{(1+\jmath^2\rho^4)^2} \Biggl[
E^2 - 8\jmath L z (E + 6 \jmath L z) + \frac{L^2}{\rho^2}
+2 \jmath^2 L^2 \rho^2 + \jmath^4 L^2 \rho^6
+ \biggl(4\sqrt{2} \jmath \rho z \frac{E + 4\jmath L z}{1+\jmath^2 \rho^4}\biggr)^2
\Biggr] \\
&\quad + \frac{\chi}{1+\jmath^2\rho^4} \,.
\end{split}
\end{equation}

\begin{figure}[h]
\captionsetup[subfigure]{labelformat=empty}
\centering
\hspace{-0.2cm}
\subfloat[\centering $m=1$, $\jmath=10^{-5}$]{{\includegraphics[width=7.5cm]{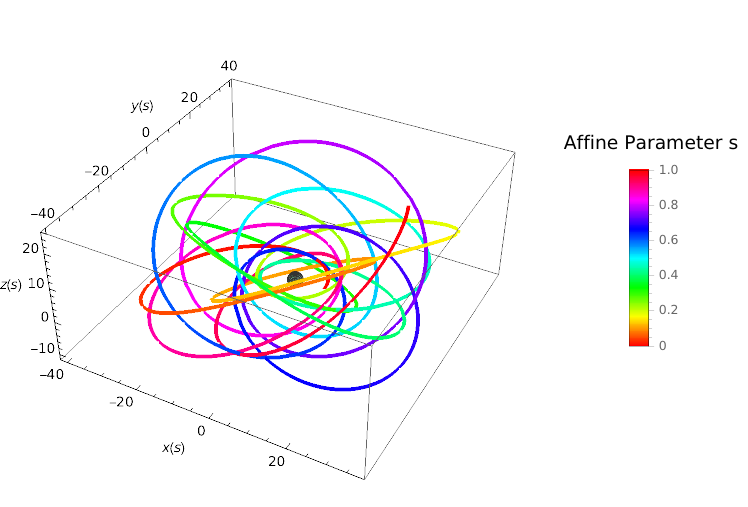}}}%
\subfloat[\centering $m=1$, $\jmath=10^{-5}$]{{\hspace{-0.3cm} \includegraphics[width=7.5cm]{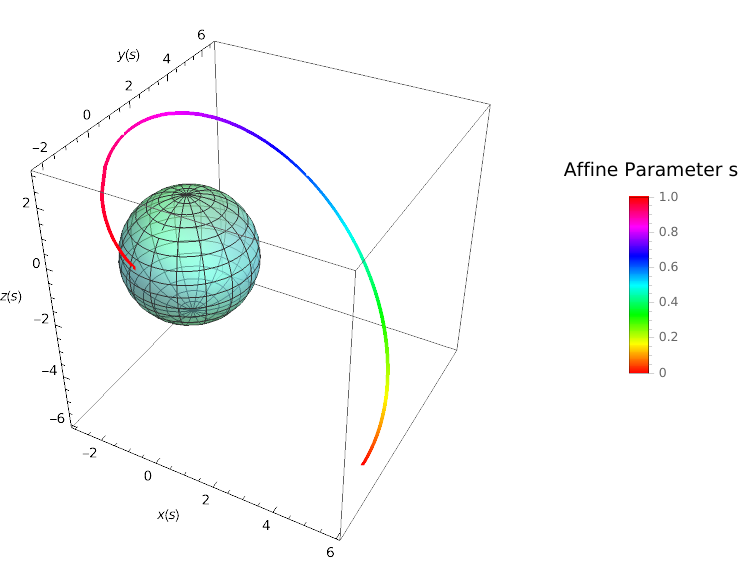}}}
\caption{\small Black hole~\eqref{swirling-bh} with two different orbits.
The left one represents the case where $\theta$(0) = $\pi/2$, $\dot{t}(0)$ = 1.30703, $s_{max} =10000$, $L = 4.24264$, $E =0.980274$.
The second one shows an unstable orbit, the test particle is attracted towards the black hole; initial parameters: $\theta$(0) = $3\pi/4$, $\dot{t}(0)$ = 1.23274, $s_{max} =43.07$, $s_{min} = 0$, $L=2.11463$, $E=0.924915$. Both picture have $r(0)=8$, $\dot{r}(0) = 0$, $t(0) = 0$, $\dot{\theta}(0) = 0 $, $\phi(0) = 0$, $\dot{\phi}(0) = \frac{3\sqrt{2}}{64}$   
}
\label{fig:unstable}
\end{figure}

\subsection{Kretschmann scalar}
\label{sec:appkre}

The complete expression of the Kretschmann scalar invariant for the Schwarzschild black hole embedded into the swirling universe is
\begin{equation}
\begin{split}
R_{\mu\nu\rho\sigma}R^{\mu\nu\rho\sigma} & = \frac{48}{r^6\big[1+\jmath^2 r^4 (1-x^2)^2\big]^6} \Biggl\{4\jmath^2 r^6 \Bigl[ -1 + 15 \jmath^2 r^4 (1-x^2)^2 -15 \jmath^4 r^8 (1-x^2)^4 + \jmath^6 r^{12} (1-x^2)^6  \Bigr] \\
&\quad + 2\jmath^2 m r^5 (1-x^2) \Bigl[ 3+ 15x^2 - 5\jmath^2 r^4 (1-x^2)^2(23+3x^2) \\
&\quad + 3 \jmath^4 r^8 (1-x^2)^4 \bigl( 43 -9x^2 + \jmath^2 r^4 (x^2-3) (1-x^2)^2\bigr) \Bigr] \\
&\quad + m^2 \Bigl[ 1+7\jmath^4 r^8 (1-x^2)^4  \Bigl( 30-40\jmath^2 r^4 (1-x^2)^2 +3\jmath^4r^8(1-x^2)^4 \Bigr) \Bigr] \Biggr\} \,.
\end{split}
\end{equation}
We clearly understand that the spacetime has a curvature singularity only at $r=0$, as the static spherically symmetric black hole seed.

\subsection{Black hole geodesics}
\label{sec:appbh}

The conserved charges equations are
\begin{subequations}
\label{geod-bh}
\begin{align}
\dot{t} & = \frac{r [ E + 4 \jmath L \cos \theta (r-2 m) ]}{(r-2m) (1 + \jmath^2 r^4 \sin^4\theta)} \,, \\
\dot{\phi} & = \frac{L - \jmath r^3 \sin^2\theta
[ 4 \cos\theta (E-8\jmath L m \cos\theta) - \jmath^3 L r^5 \sin^6\theta + 2\jmath L r (9\cos^2\theta-1) ]}{r^2 \sin^2\theta (1+\jmath^2 r^4 \sin^4\theta)} \,.
\end{align}
\end{subequations}
$L$ is the angular momentum and $E$ is the energy.
From the conservation of the four-momentum it follows
\begin{equation}
\label{4mom-BH}
\begin{split}
& \bigl( 1 + \jmath^2 r^4 \sin^4\theta \bigr)
	\biggl[-\frac{r}{r-2m}
\biggl(\frac{  E + 4 \jmath L \cos\theta (r-2m) }{1 + \jmath^2 r^4 \sin^4\theta}\biggr)^2
+ \frac{\dot{r}^2}{1-\frac{2 m}{r}}+r^2 \dot{\theta}^2\biggr] \\
&\quad + \frac{L^2 (1+\jmath^2 r^4 \sin^4\theta)^2}{r^2 \sin^2\theta}
= \chi \,.
\end{split}
\end{equation}

\section{Zipoy--Voorhees spacetime embedded in the swirling universe}
\label{app:zipoy}

We apply the procedure described in section~\ref{sec:generation} to a slightly more general metric with respect to the Schwarzschild black hole, the Zipoy--Voorhees metric.
This class of spacetime is relevant in General Relativity because, thanks to its richer multipolar expansion, can be used to model the exterior gravitational field of planets or stars.
Moreover, it can be of some interest, when supported with a conformally coupled scalar field, to build hairy black holes or wormholes such as the Bekenstein \cite{bekenstein} or the Barcelo--Visser \cite{barcelo-visser} ones, as explained in~\cite{stationary}\footnote{Actually the associated complex Ernst field equations remain the same of the pure general relativistic case, so as the main structure functions in the metric.
Only the decoupled function $\gamma$ have to be slightly modified according to~\cite{stationary}.}.
In particular, the presence of the swirling background could be useful in the wormhole configuration to improve both the stability and the traversability properties of the solution.

We start by casting the Zipoy--Voorhees seed in terms of the magnetic LWP metric~\eqref{LWP-m}, in prolate spherical coordinates
\begin{equation} 
\label{dsbarxy}
{d\bar{s}}^2 = \bar{f} \bigl( d\phi - \bar{\omega} d\tau \bigr)^2
+ \frac{1}{\bar{f}} \biggl[ -\rho^2 d\tau^2 + \kappa^2 (x^2-y^2) e^{2\bar{\gamma}}  \biggl( \frac{{dx}^2}{x^2-1} + \frac{{d y}^2}{1-y^2} \biggr) \biggr] \,,
\end{equation}
where
\begin{align}
\bar{f}(x,y) & = \kappa^2 \biggl(\frac{x-1}{x+1} \biggr)^{-\delta} (x^2-1)(1-y^2) \,, \\
 \label{gammabar}
\bar{\gamma}(x,y) & = \frac{1}{2} \log \left[ \kappa^2 \biggl(\frac{x-1}{x+1} \biggr)^{-2\delta} (x^2-1)(1-y^2) \biggl( \frac{x^2-1}{x^2-y^2} \biggr)^{\delta^2} \right] \,, \\
\rho(x,y) & = \kappa \sqrt{(x^2-1)(y^2-1)} \,. \label{rhobar}
\end{align}
Clearly this metric reduces to the static Schwarzschild black hole treated in Sec.~\ref{sec:generation} for $\delta=1$.
For generic values of $\delta \neq 1$ the metric looses the spherical symmetry and presents naked singularities outside the event horizon, hence it is not suitable for describing legitimate black holes in pure General Relativity.
However for $\delta=1/2$ it represents, when properly coupled  with a scalar field, the first hairy black hole ever discovered~\cite{bekenstein}.

Thanks to the Ehlers transformation~\eqref{sol-Er}, and following exactly the same procedure illustrated in Sec.~\ref{sec:generation} we are able to embed the Zipoy--Voorhees metric into the swirling background.
The $\bar{\gamma}$ function remains unvaried in the process, while
\begin{align}
\label{fbarbar}
\bar{f}(x,y) & =  \frac{\kappa^2 (1-y^2)(x^2-1)^{1+\delta}}{(x-1)^{2\delta}+\jmath^2(1+x)^{2\delta}(x^2-1)^2(1-y^2)^2} \,, \\
\label{omegabarbar}
\bar{\omega}(x,y) & = 4 \jmath \kappa^2 y (x-\delta) + \omega_0 \,.
\end{align}
The metric defined by~\eqref{dsbarxy} and~\eqref{gammabar},~\eqref{omegabarbar} represents the $\delta$ extension of the spacetime~\eqref{bh-rot-universe}, therefore the Zipoy--Voorhees spacetime immersed in the rotating background described in Sec.~\ref{sec:analysis-background}.
Further generalisations with angular momentum can be built straightforwardly, starting with seeds of the family of the Tomimatsu--Sato solutions~\cite{cosgrove,hkx}.

\end{document}